# Glassy disordered ground states in the frustrated pyrochlore and fluorite antiferromagnets NaCd$M_2$F$_7$ ($M$ = Ni$^{2+}$, Mn$^{2+}$)

Andrej Kancko,[1] Cinthia Antunes Corrêa,[2] and Ross Harvey Colman [1*]

[1] *Charles University, Faculty of Mathematics and Physics, Department of Condensed Matter Physics, Ke Karlovu 5, 121 16 Prague 2, Czech Republic*
[2] *Institute of Physics of the Czech Academy of Sciences, Na Slovance 2, Prague 8, 182 21, Czech Republic*

*Corresponding author: ross.colman@matfyz.cuni.cz*

**Abstract**

We report the crystal structures, magnetic and thermodynamic properties of two magnetically frustrated $A'A''M_2$F$_7$ -type antiferromagnets, NaCdNi$_2$F$_7$ and NaCdMn$_2$F$_7$. While NaCdNi$_2$F$_7$ forms a stable pyrochlore structure (SG: *Fd-3m*, #227) with magnetic $S$ = 1 Ni$^{2+}$ ions on the frustrated pyrochlore 16$c$ site and fully disordered non-magnetic Na$^+$/Cd$^{2+}$ ions on the pyrochlore 16$d$ site, NaCdMn$_2$F$_7$ favors the *defect*-fluorite structure (SG: *Fm-3m*, #225) with magnetic $S$ = 5/2 Mn$^{2+}$ and non-magnetic Na$^+$ and Cd$^{2+}$ ions fully disordered on the fluorite 4$a$ site. This is a result of the Mn$^{2+}$ ionic radius being too close to the average Na$^+$/Cd$^{2+}$ ionic radius, hindering the cationic site ordering towards the stable pyrochlore structure. In both cases, dominant antiferromagnetic interactions θ$_{CW,Ni}$ = -91.2(5) K and θ$_{CW,Mn}$ = -42.4(4) K are noted, with no magnetic transition until $T_{f,Ni}$ = 3.2 K and $T_{f,Mn}$ = 2.0 K, implying substantial frustration, with frustration indices $f_{Ni}$ = 28.5 and $f_{Mn}$ = 21. AC susceptibility measurements and bifurcation of ZFC/FC low-field magnetization indicate a spin-glass-like ground-state, precipitated by the magnetic-bond-disorder stemming from the inherent structural disorder.

## Introduction

The pyrochlore lattice, a 3D network of corner-sharing tetrahedra, is a prototypical structure for three-dimensional geometrical frustration. The specific vertex-sharing tetrahedral arrangement of spins leads to competing magnetic exchange interactions that prevent the system from achieving a trivial magnetic order. [1,2] The frustration frequently gives rise to highly-degenerate exotic ground-states, including spin glasses, spin ices, spin liquids, topological states, or even complex magnetic orders. [3–6] Materials that realise these frustrated magnetic lattices provide mechanisms to test and refine the predictions of these increasingly complex theoretical models of the interactions of electrons in solids.

There are two noteworthy families of compounds that contain the frustrated pyrochlore lattice. The main representatives are the $A_2B_2X_6X'$ pyrochlores ($A$, $B$ = rare earth or transition metals; $X$ = O, F and $X'$ = O, F, S). [7–10] In this family, both the $A$ and $B$ sites each contain one independent sublattice of corner-sharing tetrahedra. The other family, the $AB_2X_4$ spinels ($B$ = rare earth, $X$ = S, Se) only contain a single pyrochlore sublattice on the rare-earth $B$-site. [11,12] Deviations from the ideal pyrochlore structure can also be found in the so-called "β-pyrochlores" $AM^{2+}M^{3+}$F$_6$ ($A^+$ = Rb, Cs; $M^{2+}$ = Mn, Co, Ni and $M^{3+}$ = V, Cr, Fe) [13] or "breathing pyrochlores" in chromium-based $AA'$Cr$_4$X$_8$ spinels ($X$ = O, S, Se). While the β-pyrochlores contain disordered mixed-valence magnetic ions $M^{2+}$ and $M^{3+}$ on the pyrochlore $B$-site and vacancies on the $A$-site, the breathing pyrochlores contain a network of alternating small and large tetrahedra. [14–16]

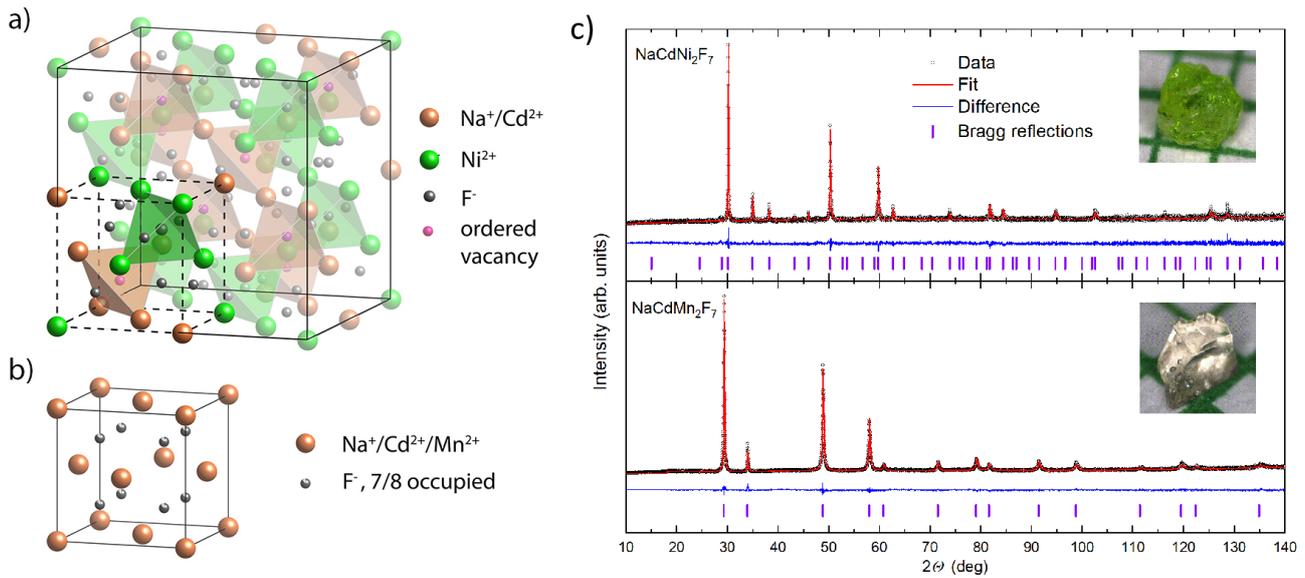

Figure 1. (a) The pyrochlore unit cell of NaCdNi$_2$F$_7$, with the (b) disordered defect fluorite cell of NaCdMn$_2$F$_7$. The cation ordering and vacancy ordering of the pyrochlore structure leads to a doubling of the cell along each axis. We have highlighted the representative comparison between the two structures as a dashed sub-cell within the pyrochlore structure. Ionic sizes are not-to-scale. (c) Shows powder diffraction data with Rietveld refinement fits and residuals for both compounds, as well as photos of the resulting single-crystal fragments on mm-square paper, as insets.

In the past couple of decades, the rare-earth-based $A_2B_2O_7$ pyrochlore oxides have been extensively investigated due to the availability of large single crystals and the observation of a plethora of exotic magnetic ground-states. [17] These ranged from unconventional spin-glass (Y$_2$Mo$_2$O$_7$) [18], to ordered (Tb$_2$Sn$_2$O$_7$) and disordered (Ho$_2$Ti$_2$O$_7$) spin-ice [19,20], spin-liquid (Tb$_2$Ti$_2$O$_7$) [21] or order-by-disorder states (Er$_2$Ti$_2$O$_7$). [22] The complex magnetism has been found robust even against structural disorder at the pyrochlore stability boundary, when the size difference between $A$ and $B$ site ions is too small and a defect fluorite structure is instead formed. [23,24]

The variety of properties seen in the 4$f$ rare-earth pyrochlores comes about due to the enhanced importance of single-ion anisotropy as a result of a large spin-orbit coupling, as well as interaction with neighbors via crystal-electric-field effects or further-neighbor exchange. The weak dipolar coupling between the magnetic 4$f$ orbitals typically leads to low interaction strengths ($|\theta_{CW}| \sim 10^0 – 10^1$ K), requiring very low temperatures to study the exotic ground-states. Implementing a late 3$d$-metal magnetic ion (from Fe$^{2+}$ to Cu$^{2+}$) with much greater coupling strengths thanks to superexchange through the more extensive 3$d$ orbitals ($|\theta_{CW}| \sim 10^1 – 10^2$ K) would solve this issue, allowing the study of magnetic properties at more accessible temperatures. However, charge balance and pyrochlore structure stability constraints for the $A_2B_2O_7$ oxides hinder this possibility, only allowing the implementation of higher-spin-state transition metals (4+ or 5+) on the pyrochlore $B$-site. [25]

The family of A'A"B$_2$F$_7$ pyrochlore fluorides offers a solution. Because of the similar ionic radii of fluorine ($r_{F^-}$ = 1.33 Å) and oxygen ($r_{O^{2-}}$ = 1.4 Å), as well as similar electronegativities ($X_F$ = 3.98 and $X_O$ = 3.44), this material family maintains a stable pyrochlore structure by replacing the divalent oxygen (O$^{2-}$) by a monovalent fluorine (F$^{1-}$). The smaller charge of fluorine then allows for a magnetic 3d-metal ion with a low oxidation state (2+) on the pyrochlore B-site. Due to charge balancing, however, there is a necessity for a fully mixed occupancy of the $A$-site by a monovalent $A'^+$ and a divalent $A''^{2+}$ cation, $(A'^+_{0.5}A''^{2+}_{0.5})_2B^{2+}_2F^{1-}_7$, yielding an effective $A^{1.5+}$ oxidation state to maintain the 2:2:7 pyrochlore stoichiometry, ($A^{1.5+}_2 B^{2+}_2 F^{1-}_7$). This introduces chemical disorder, which in turn locally influences the magnetic exchange pathways due to the ionic size mismatch between the A'$^{+}$ and A"$^{2+}$ cations. This weak magnetic bond disorder in the Heisenberg pyrochlore antiferromagnet then typically precipitates a frozen spin-glass ground-state. [26,27]

Beyond the initial synthesis and structural report of some polycrystalline $A'A"M_2F_7$ pyrochlore fluorides in 1970 [28], their magnetism has been neglected until the first successful floating-zone growths of large Na$A"M_2F_7$ single crystals (with $A"^{2+}$ = Ca, Sr, Cd; and $M^{2+}$ = Co, Ni, Fe, Mn). [29–33] All members exhibit strong antiferromagnetic interactions $|\theta_{CW}| \sim 70 - 140$ K with no apparent magnetic transition until $T_f \sim 2 - 4$ K, indicating a sizable frustration with the frustration index $f = \frac{|\theta_{CW}|}{T_f}$ ranging between 19 – 58. In all reported cases, the materials behave as thermal spin liquids until the frustration is quenched by means of spin freezing into a disordered glassy ground-state, selected from the large manifold of degenerate antiferromagnetic states. AC susceptibility measurements suggest a spin-glass ground-state in NaCaFe$_2$F$_7$, NaSrFe$_2$F$_7$ and NaSrMn$_2$F$_7$ [33], in agreement with the ab initio DFT + U + SOC calculations of magnetic and critical properties of NaSrMn$_2$F$_7$. [34] Inelastic neutron scattering and NMR studies of the $J_{eff}$ = ½ Na$A"$Co$_2$F$_7$ ($A"$ = Ca, Sr) suggest a glass-like short-range-ordered state with antiferromagnetic XY-type clusters, while ESR measurements indicate a coexistence of two distinct magnetic phases - a cooperative paramagnetic phase with a gapless excitation mode, and a gapped spin-glass phase below 20 K. [35–38] A recent study of the isostructural $J_{eff}$ = ½ NaCdCo$_2$F$_7$ showed a positive correlation between the freezing temperature, $T_f$ and the Na$^+$/$A"^{2+}$ ionic-size-mismatch, comparing with the isostructural Na$A"$Co$_2$F$_7$ members ($A"$ = Ca, Sr). [31] In spite of the spin freezing also seen in the $S$ = 1 NaCaNi$_2$F$_7$, persistent spin dynamics down to 75 mK were detected via muon spin relaxation spectroscopy. [39] Inelastic neutron scattering also revealed a continuum of magnetic scattering with low-energy pinch-points, suggesting that NaCaNi$_2$F$_7$ shows important features of a quantum spin liquid ground-state in the spin-1 antiferromagnetic Heisenberg model on a pyrochlore lattice. [40,41]

In this paper, we report the successful synthesis and structural, magnetic, and thermodynamic characterisation of two new additions to the $A'A"M_2F_7$ fluoride family: $S$ = 1 pyrochlore fluoride antiferromagnet NaCdNi$_2$F$_7$, and $S$ = 5/2 defect-fluorite antiferromagnet NaCdMn$_2$F$_7$.

**Experimental**

Single crystals of NaCdNi$_2$F$_7$ and NaCdMn$_2$F$_7$ were grown by slow crystallization from the melt in platinum crucibles sealed in an argon atmosphere. A stoichiometric mix of dry precursor binary fluorides was prepared and ground inside a glove-box, filled into a platinum crucible, and crimp-sealed in argon to prevent the presence of air and moisture, avoiding possible oxidation. The crucibles were heated inside a box furnace, with the heating protocols programmed to reach 850°C (NaCdMn$_2$F$_7$) and 950°C (NaCdNi$_2$F$_7$) at the rate of 2°C/min, dwell there for 3 hours, then slow-cool by 200°C over 7 days and finally rapidly cool to room temperature. This was repeated 4 times, with intermittent grinding to improve homogeneity and ensure a complete reaction. A large multi-grain crystal (∼ 0.5 g) of NaCdMn$_2$F$_7$ was obtained, which could be broken into smaller (∼ 30 mg) single-grain pieces. However, only several small (∼ 1-2 mg) single-grain crystals of NaCdNi$_2$F$_7$ could be obtained, while the rest of the reacted material consisted of sintered polycrystalline powder. Single-grain pieces were used for structural solution via single crystal X-ray diffraction, although only NaCdMn$_2$F$_7$ crystals were big enough to be oriented using a PhotoScience Laue diffractometer to perform directional magnetic and thermodynamic property measurements.

Single-crystal X-ray diffraction experiments were performed at 95 K on a Rigaku SuperNova diffractometer, using a mirror collimated Mo K$\alpha$ (λ = 0.7173 Å) radiation from a microfocus sealed tube, and an Atlas S2 CCD detector. Diffraction data were integrated using CrysAlis Pro [42] with an empirical absorption correction using spherical harmonics (NaCdMn$_2$F$_7$), and an empirical absorption correction using spherical harmonics combined with a numeric absorption correction based on Gaussian integration over a multifaceted crystal model (NaCdNi$_2$F$_7$). The structure was solved by charge flipping using the program Superflip [43] and refined by full-matrix least squares on F$^2$ in Jana2020. [44] Structural graphics were created using Jana2020 and Vesta. [45]

To confirm that the single crystal structure solutions were representative of the bulk of the sample, a small number of individual grains were selected, crushed into a fine powder and analyzed by Powder X-Ray Diffraction (PXRD) at room temperature ($T$ = 300 K) using a Bruker D8 Advance diffractometer with Cu Kα radiation (λ = 1.5418 Å). Structural Rietveld refinement was performed using Topas Academic V6. [46]

Temperature and field-dependent magnetization measurements on oriented crystals of NaCdMn$_2$F$_7$ and polycrystalline NaCdNi$_2$F$_7$ were performed in a Quantum Design Magnetic Property Measurement System (model MPMS-XL 7T), using the DC extraction method (NaCdMn$_2$F$_7$) as well as the reciprocating sample option (RSO, NaCdNi$_2$F$_7$). Spin dynamics were probed by temperature and frequency dependent AC susceptibility measurements using the ACMS II option in a Quantum Design Physical Property Measurement System (PPMS). Specific heat between 1.8 – 300 K was measured inside the PPMS using the relaxation method, with the addition of a $^3$He insert for low-temperature measurements down to 0.4 K. A Quantum Design heat capacity puck with a non-magnetic sapphire stage was used, with Apiezon-N grease for mounting the crystal. Measurement of a non-magnetic analogue NaCdZn$_2$F$_7$ was used in the subtraction of the phonon contribution to the specific heat of NaCdNi$_2$F$_7$ and NaCdMn$_2$F$_7$ to estimate the magnetic specific heat.

**Results and Discussion**

*Structure analysis*

Single-crystal experimental results follow in Tables 1-3.

The crystal structure of NaCdNi$_2$F$_7$ obtained by SCXRD at 95 K is cubic, space group *Fd-3m* (#227), $Z$ = 8, and with unit cell parameter $a_{Ni}$ =10.2245(10) Å. This is consistent with the expected pyrochlore structure, shown in Figure 1 (a). An initial model with Na and Cd sharing the 16*d* site with a Na:Cd ratio of 1:1 gave $R_{obs}$ = 3.51% and $GOF_{obs}$ = 4.49%. However, refined F2 and Na:Cd occupancies, whilst maintaining a full site occupancy of the 16*d* site, lowered the residue values to $R_{obs}$ = 1.83% and $GOF_{obs}$ = 1.19%. This gives the final chemical composition Na$_{0.892}$Cd$_{1.108}$Ni$_2$F$_{6.982}$. An attempt at refining only the Na/Cd occupancy site gave a chemical composition of Na$_{0.894}$Cd$_{1.106}$Ni$_2$F$_7$, making almost no difference in the residue values, slightly increasing the $GOF_{obs}$ from 1.19% to 1.23%. The error on the F2 site occupancy is in the third position [0.982(9)], while the Na/Cd occupancies variations are within the error [0.4459(19) for Na and 0.5541(19) for Cd], which suggests that the refinement is much less sensitive to the light element F2 site occupancy. Charge balancing arguments, however, would expect that an increase in the Cd$^{2+}$ cation occupancy above the ideal 1:1:2:7 Na:Cd:Ni:F ratios, and at the expense of the Na$^+$ occupancy, would require an increase in the total F$^-$ anion stoichiometry. We saw no evidence for additional F- sites within the structure to charge balance the slight excess of the Cd$^{2+}$ cation, but expect that the minor occupation of interstitial or vacancy sites necessary for this are below the resolution of our measurements.

The crystal structure of NaCdMn$_2$F$_7$ obtained by SCXRD at 95 K is cubic, space group *Fm-3m* (#225), $Z$ = 1, and with unit cell parameter $a_{Mn}$ = 5.2461(5) Å. The cation site (4*a*) is shared by Na$^+$, Cd$^{2+}$ and Mn$^{2+}$ in a ¼ : ¼ : ½ ratio, remaining fully disordered, as seen in Figure 1 (b). The ADPs of Na/Cd/Mn were kept fixed to be the same for all atoms sharing the site, and occupancies were fixed to the ideal ratios. The occupancy of fluorine on the 8*c* site was constrained to 7/8 to retain the expected NaCdMn$_2$F$_7$ stoichiometry, which lead to a residue value of $R_{obs}$ = 1.97% and $GOF_{obs}$ = 1.97%. This solution is consistent with the defect-fluorite structure, where F is known to have a disordered vacancy, decreasing its occupancy from 8 to 7 atoms per unit cell, and analogous to oxides of the same type. [24,47] In contrast, assuming full F site occupancy (NaCdMn$_2$F$_8$) lead the refinement to significantly worse residue values $R_{obs}$ = 2.87% and $GOF_{obs}$ = 3.90%, as expected from the initial fluorine content in the stoichiometric mix of precursor binary fluorides.

The powder X-ray diffraction patterns of NaCdNi$_2$F$_7$ and NaCdMn$_2$F$_7$ confirm the expected pyrochlore (SG: *Fd-3m*, #227:2) and defect-fluorite (SG: *Fm-3m*, #225) structures, respectively. The refined lattice parameters, $a_{Ni}$ = 10.257(1) Å and $a_{Mn}$ = 5.2714(2) Å, are in line with previous reports [28] and consistent with our SCXRD results (taken at 95 K) after considering thermal expansion.

Comparing the PXRD patterns of the two samples, seen in Figure 1 (c), highlights the structural differences, where superstructure reflections stemming from the cation ordering and fluoride ion vacancy ordering of the pyrochlore lattice are clearly present in NaCdNi$_2$F$_7$ *but absent in* NaCdMn$_2$F$_7$. The reason behind NaCdMn$_2$F$_7$ crystallizing in the defect-fluorite structure instead of the pyrochlore structure lies in the large ionic radius of Mn$^{2+}$, which in the high-spin state and a six-fold coordination amounts to $r_{Mn2+}$ = 0.83 Å. [48] The simplest and most-widely used descriptor of the $A_2B_2O_7$ pyrochlore structure stability region is given by the ratio of the *A*-site and *B*-site ionic radii $RR = R_A/R_B$, where a stable pyrochlore structure on the oxide series is expected for RR ∈ [1.46, 1.80]. [8] Using 8-fold-coordinated Shannon radii [48] for the average ionic radius of the *A*-site shared by Na$^+$ (1.18 Å) and Cd$^{2+}$ (1.10 Å), gives $R_A$ = 1.14 Å and $RR$ = 1.37, outside the lower bound for a stable pyrochlore structure. This results in full cation disorder, as well as disorder of the vacancy of the 7/8$^{th}$ occupied fluoride site in the fluorite structure. In contrast, $RR$ = 1.65 in NaCdNi$_2$F$_7$ lies well in the middle of the stability region, resulting in a stable pyrochlore structure.

Table 1: Single crystal experimental results and structural refinement of NaCdMn$_2$F$_7$ and NaCdNi$_2$F$_7$

| Crystal data | Mn | Ni |
|---|---|---|
| Chemical formula | NaCdMn$_2$F$_7$ | Na$_{0.892}$Cd$_{1.108}$Ni$_2$F$_{6.982}$ |
| Crystal system, Space group | cubic, Fm-3m (#225) | cubic, Fd-3m (#227): setting 2 |
| Temperature (K) | 95 | 95 |
| $a$ (Å) | 5.2461(5) | 10.2245(10) |
| Crystal size (mm) | 0.119 x 0.087 x 0.017 | 0.22 x 0.16 x 0.105 |
| Color, shape | light yellow, plate | lime green, irregular |
| $V$ (Å$^3$), $Z$ | 144.38(2), 1 | 1068.88(18), 8 |
| $\lambda$ (Mo$K\alpha$) (Å) | 0.71073 | |
| $\mu$ (mm$^{-1}$) | 8.101 | 11.483 |
| $\vartheta$ Range (°) | 6.74, 37.29 | 3.45, 29.68 |
| $F(000) / D_x$ (Mg.m$^{-3}$) | 172/4.3504 | 1455, 4.9102 |
| Index ranges | -8 ≤ $h$ ≤ 8<br>-8 ≤ $k$ ≤ 8<br>-8 ≤ $l$ ≤ 8 | -14 ≤ h ≤ 13<br>-13 ≤ k ≤ 13<br>-14 ≤ l ≤ 14 |
| independent reflections | 35 | 95 |
| measured reflections | 1189 | 4405 |
| Observed data ($I > 3\sigma(I)$) | 35 | 74 |
| Refinement method | Refinement on $F^2$ | |
| $R_{int}$ | 0.0447 | 0.0949 |
| Diffractometer | SuperNova, Dual, Mo at home/near, AtlasS2 | |
| Absorption correction | Multi-scan<br>CrysAlisPro 1.171.41.117a (Rigaku Oxford Diffraction, 2021) Empirical absorption correction using spherical harmonics, implemented in SCALE3 ABSPACK scaling algorithm. | |
| Structure solution | Charge flipping | |
| Refinement Software | Jana2020 | |
| $R[F^2 > 3\sigma(F^2)]$, $wR_2(F^2)$, GOF | 0.0197, 0.0775, 1.9723 | 0.0183, 0.0534, 1.1396 |
| No. of Parameters, restraints, constraints | 4, 0, 3 | 13, 0, 2 |
| $\Delta\rho_{max}$, $\Delta\rho_{min}$ (e Å$^{-3}$) | 0.48, -0.62 | 0.34, -0.49 |
| CCDC Number | 2389766 | 2389764 |

Table 2: Atomic positions and anisotropic thermal displacement parameters of $NaCdMn_2F_7$

| | | | | | | | $U_{ij}$ (×10$^4$ Å$^2$) | | | | | | |
|---|---|---|---|---|---|---|---|---|---|---|---|---|---|
| Atom | Site | x | y | z | Occ. | Site sym. | $U_{11}$ | $U_{22}$ | $U_{33}$ | $U_{12}$ | $U_{13}$ | $U_{23}$ | $U_{ani}$ |
| | | | | | | NaCdMn$_2$F$_7$ | | | | | | | |
| Na | 4a | 0 | 0 | 0 | ¼ | m-3m | 232(6) | 232(6) | 232(6) | 0 | 0 | 0 | 232(3) |
| Cd | 4a | 0 | 0 | 0 | ¼ | m-3m | 232(6) | 232(6) | 232(6) | 0 | 0 | 0 | 232(3) |
| Mn | 4a | 0 | 0 | 0 | ½ | m-3m | 232(6) | 232(6) | 232(6) | 0 | 0 | 0 | 232(3) |
| F | 8c | ¼ | ¼ | ¼ | ⅞ | -43m | 560(2) | 560(2) | 560(2) | 0 | 0 | 0 | 558(14) |

Table 3: Atomic positions and anisotropic thermal displacement parameters of $NaCdNi_2F_7$

| | | | | | | | $U_{ij}$ (×10$^4$ Å$^2$) | | | | | | |
|---|---|---|---|---|---|---|---|---|---|---|---|---|---|
| Atom | Site | x | y | z | Occ. | Site sym. | $U_{11}$ | $U_{22}$ | $U_{33}$ | $U_{12}$ | $U_{13}$ | $U_{23}$ | $U_{ani}$ |
| | | | | | | Na$_{0.892}$Cd$_{1.108}$Ni$_2$F$_{6.982}$ | | | | | | | |
| Na | 16d | ½ | ½ | ½ | 0.4462(17) | .-3m | 93(4) | 93(4) | 93(4) | -7(9) | -7(9) | -7(9) | 93(2) |
| Cd | 16d | ½ | ½ | ½ | 0.5538(17) | .-3m | 93(4) | 93(4) | 93(4) | -7(9) | -7(9) | -7(9) | 93(2) |
| Ni | 16c | 0 | 0 | 0 | 1 | .-3m | 56(4) | 56(4) | 56(4) | 0(1) | 0(1) | 0(1) | 56(2) |
| F1 | 48f | 0.33201(15) | ⅛ | ⅛ | 1 | 2.mm | 105(5) | 131(9) | 105(5) | 0 | 33(5) | 0 | 113(4) |
| F2 | 8b | ⅜ | ⅜ | ⅜ | 0.982(9) | -43m | 125(9) | 125(9) | 125(9) | 0 | 0 | 0 | 125(5) |

Magnetic susceptibility

NaCdNi$_2$F$_7$ DC susceptibility

Temperature and field-dependent DC magnetization $M(T,H)$ measurements were conducted on polycrystalline NaCdNi$_2$F$_7$. The temperature dependence of magnetization $M(T)$ in a constant field of $\mu_0 H$ = 2 T is shown as $\chi(T)$ and $\chi(T)^{-1}$ for NaCdNi$_2$F$_7$ in Figure 2. The field-dependent magnetization measured at 2 K is presented in the lower inset of Figure 2 and shows a linear response up to $\mu_0 H$ = 7 T, without hysteresis effects, justifying the consideration of the magnetisation data as DC susceptibility as $\chi(T) = M(T)/H$. At $\mu_0 H$ = 7 T the moment remains far below the expected saturated moment (2 $\mu_B$) for an orbitally quenched moment on $S$ = 1 Ni$^{2+}$. A Curie-Weiss fitting was performed in the 100 – 300 K range, resulting in a Curie-Weiss temperature $\theta_{CW}$ = -91.2(5) K and the effective moment $\mu_{eff}$ = 3.15(1) $\mu_B$/Ni$^{2+}$. In spite of the strong antiferromagnetic interactions evidenced by the large negative Curie-Weiss temperature, no features in the susceptibility are seen until cooling below $T_f$ = 3.2 K. The upper inset of Figure 2 shows a low-field measurement of susceptibility, where a sharp cusp indicates a magnetic transition that yields a significant Ramirez frustration index of $f = -\theta_{CW}/T_f$ = 28.5. [1] A bifurcation of the ZFC/FC data at the transition is characteristic of either magnetic order or spin-glass-like freezing, as seen with the spin-glass state of the previously-reported pyrochlore fluorides. [29–33] The glassy nature of this transition in NaCdNi$_2$F$_7$ is confirmed by AC susceptibility and specific heat measurements, discussed later. From the electron configuration of $S$ = 1 Ni$^{2+}$ ($t_{2g}^6 e_g^2$) in a six-fold-coordinated octahedral environment, where the three $t_{2g}$ shells are fully occupied and the two $e_g$ shells are singly occupied, one would expect a quenching of the orbital moment, leading to a spin-only effective moment of 2.83 $\mu_B$/Ni$^{2+}$. The larger effective moment extracted from the Curie-Weiss fit, however, suggests a small but non-zero orbital contribution, likely coming about from the crystal field effects from the trigonally-distorted NiF$_6$ octahedra. This effect, although stronger, was also seen in the isostructural analogue NaCaNi$_2$F$_7$ ($\mu_{eff}$ = 3.7 $\mu_B$). [32] Table 4 shows a compilation of the magnetic properties for all members of this transition metal fluoride family.

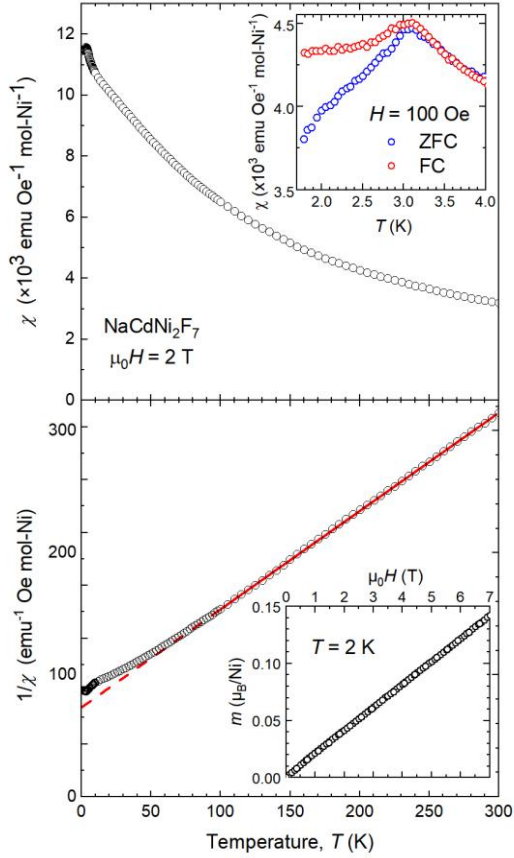

Figure 2. Temperature dependent magnetic susceptibility (upper) and inverse susceptibility (lower) data for NaCdNi$_2$F$_7$. The red solid (dashed) line is the Curie-Weiss fit (extrapolation). Upper inset is a low-field measurement, indicating the ZFC-FC history dependence below $T_f$ = 3.2 K, whilst the lower inset highlights the linear magnetisation that is seen at all temperatures.

## NaCdMn$_2$F$_7$ DC susceptibility

The magnetic susceptibility and inverse susceptibility of NaCdMn$_2$F$_7$, measured along $H \parallel$ [100], [110] and [111] directions in an $H$ = 2000 Oe field are shown in Figure 3. The isothermal magnetization $M(H)$ measured at 2 K, shown in the lower inset of Figure 3, displays a subtle curvature up to $\mu_0 H$ = 7 T, but is linear in the vicinity of $H$ = 2000 Oe and shows no hysteresis upon decreasing the field, again supporting our interpretation of DC susceptibility as $\chi = M/H$. Due to the strong magnetic frustration, the magnetization at 7 T reaches only around 20% of the saturated value expected for an orbitally quenched $S$ = 5/2 Mn$^{2+}$ ion (5 $\mu_B$), which is typical in compounds of 3$d$ transition metal ions as a result of the crystal-field splitting. The $M(H)$ and $\chi(T)$ curves overlay in all high-symmetry directions, suggesting a Heisenberg (isotropic) nature of this defect-fluorite antiferromagnet. The inverse susceptibility data along [111] were fitted to the Curie-Weiss law in the 100 – 300 K range, yielding a Curie-Weiss temperature of $\theta_{CW}$ = -42.4(4) K and an effective moment of $\mu_{eff}$ = 5.88(2) $\mu_B$/Mn$^{2+}$. The negative Curie-Weiss temperature confirms dominant antiferromagnetic interactions, while the magnitude is ∼ 2 times lower than that observed in the pyrochlore analogue NaSrMn$_2$F$_7$. [33] The crossover from ordered pyrochlore structure to disordered defect-fluorite is expected to result in significant changes to the magnetic behaviour. The magnetic ions no longer reside on a pyrochlore lattice, but instead now sit on an FCC lattice with significant (50 %) dilution. The FCC lattice is also considered geometrically frustrated, however, to a much lesser degree than the pyrochlore lattice due to a reduced number of soft modes in the correlated paramagnetic state. [49] The local disorder that the significant dilution brings about results in a weaker but still sizable mean-field interaction. Despite these strong exchange interactions, no magnetic transition is apparent down to approximately $T_f$ = 2.0 K, where a rounded cusp in $\chi$(T) begins to appear, yielding a frustration index of $f$ = 21.2. Due to the close proximity of the transition to the MPMS's base temperature (1.8 K), only a weak ZFC/FC splitting in 30 Oe field could be seen at the transition (upper inset of Figure 3). In the octahedral crystal field environment, the 3d$^5$ Mn$^{2+}$ ion usually takes the high-spin (t$_{2g}^3$e$_g^2$) electronic configuration, where each of the individual t$_{2g}$ (|xy>, |xz>, |yz>) and e$_g$ (|x$^2$-y$^2$>, |z$^2$>) orbitals is singly occupied. The spin-only effective moment expected for a free $S$ = 5/2 Mn$^{2+}$ ion (ground term $^6$S$_{5/2}$, $L$ = 0) is 5.92 $\mu_B$, which is in line with the magnitude of the measured effective moment. This result is in contrast with the effective moment seen in the NaSrMn$_2$F$_7$ pyrochlore, where some orbital contribution was seen due to crystal field effects (6.25 $\mu_B$/Mn$^{2+}$), seen in Table 4. [33]

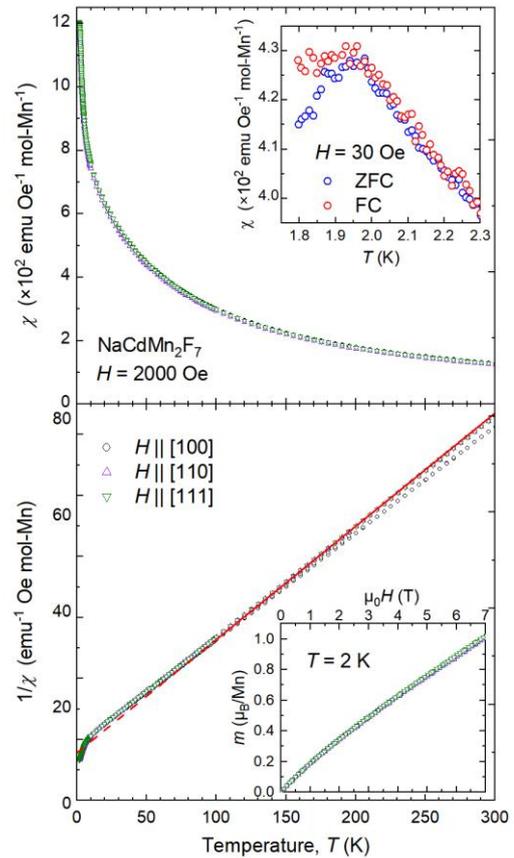

Figure 3. Temperature dependent magnetic susceptibility (upper) and inverse susceptibility (lower) data for NaCdMn$_2$F$_7$ along selected high-symmetry crystal directions. The solid (dashed) line is the Curie-Weiss fit (extrapolation). Upper inset is a low-field measurement, indicating the ZFC-FC history dependence below $T_f$ = 2.0 K, whilst the lower inset highlights the slight curvature of the magnetisation at low temperature.

Table 4: Magnetic properties comparison of previously studied AA'B$_2$F$_7$ pyrochlore (defect-fluorite) antiferromagnets.

| Compound | Space Group | S | $\mu_{eff}$ ($\mu_B$/M$^{2+}$) | $\theta_{CW}$ (K) | $T_f$ (K) | $f = \dfrac{-\theta_{CW}}{T_f}$ | References |
|---|---|---|---|---|---|---|---|
| NaCdNi$_2$F$_7$ | $Fd\bar{3}m$ | 1 | 3.15(1) | -91.2(5) | 3.2 | 28.5 | This work |
| NaCaNi$_2$F$_7$ | $Fd\bar{3}m$ | 1 | 3.7(1) | -129(1) | 3.6 | 36 | [32] |
| NaCdMn$_2$F$_7$ | $Fm\bar{3}m$ | $\dfrac{5}{2}$ | 5.88(2) | -42.4(4) | 2.0 | 21 | This work |
| NaSrMn$_2$F$_7$ | $Fd\bar{3}m$ | $\dfrac{5}{2}$ | 6.25 | -89.72 | 2.5 | 36 | [33] |
| NaCdCo$_2$F$_7$ | $Fd\bar{3}m$ | $\dfrac{3}{2}$ ($J_{eff}$ = ½) | 5.4(1) | -108(1) | 4.0 | 27 | [31] |
| NaCaCo$_2$F$_7$ | $Fd\bar{3}m$ | $\dfrac{3}{2}$ ($J_{eff}$ = ½) | 6.1(1) | -139(1) | 2.4 | 58 | [29] |
| NaSrCo$_2$F$_7$ | $Fd\bar{3}m$ | $\dfrac{3}{2}$ ($J_{eff}$ = ½) | 5.9(1) | -127(1) | 3.0 | 42 | [30] |
| NaCaFe$_2$F$_7$ | $Fd\bar{3}m$ | 2 | 5.61 | -72.79 | 3.9 | 19 | [33] |
| NaSrFe$_2$F$_7$ | $Fd\bar{3}m$ | 2 | 5.94 | -98.12 | 3.7 | 26.5 | [33] |

To directly compare the susceptibility data of different magnetic materials, one can rearrange the Curie-Weiss law into a normalized dimensionless form $C/\chi|\theta_{CW}|$ = $T/|\theta_{CW}|$ + 1. Here, the susceptibility is scaled by the effective moment ($C \sim \mu_{eff}^2$) as well as the mean-field interaction strength ($\theta_{CW} \sim J$), and is plotted against the normalized temperature $T/|\theta_{CW}|$. Considering an ideal unfrustrated antiferromagnet, the normalized high-temperature susceptibility would follow the line $y = x + 1$, with the gradient as well as y-intercept both equal to one. A long-range ordering transition would be expected to be seen as a cusp in the susceptibility in the vicinity of $T_N/|\theta_{CW}| \sim 1$. The magnetic frustration, however, pushes the magnetic transition to much lower temperature than the mean-field interaction strength $T_f/|\theta_{CW}| \ll 1$. [29–33] The positive or negative deviations from the ideal Curie-Weiss behaviour signify the onset of enhanced antiferromagnetic or ferromagnetic short-range correlations, respectively.

In Figure 4, we compare the NaCdMn$_2$F$_7$ and NaCdNi$_2$F$_7$ Curie-Weiss data, including the recently-studied NaCdCo$_2$F$_7$ [31]. All three materials follow the ideal Curie-Weiss behaviour very closely at high temperatures ($T_f/|\theta_{CW}| \geq 1$). However, significant deviations begin to appear as we approach $T_f$. NaCdMn$_2$F$_7$ starts to positively deviate from the ideal Curie-Weiss line below ~ 3.5 $T/|\theta_{CW}|$ as it develops enhanced antiferromagnetic correlations. As we approach the spin-freezing transition, however, significant ferromagnetic fluctuations begin to develop below ~ 0.1 $T/|\theta_{CW}|$. NaCdNi$_2$F$_7$ follows the ideal Curie-Weiss behaviour down to ~ 1 $T/|\theta_{CW}|$, where antiferromagnetic fluctuations begin to appear as increasing positive deviations from the ideal line approaching $T_f$ from above. For comparison, NaCdCo$_2$F$_7$ appears to follow the ideal Curie-Weiss form down to an extremely low temperature of ~ 0.06 $T/|\theta_{CW}|$ and only deviates ferromagnetically below this normalized temperature as it approaches the freezing transition.

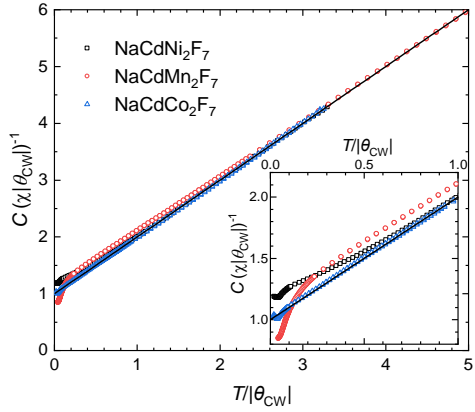

Figure 4. Rearranged dimensionless Curie-Weiss data for the known NaCd$M_2$F$_7$ ($M$ = Ni, Mn, Co) compounds. The solid line at $C/(\chi|\theta_{CW}|) = T/|\theta_{CW}| + 1$ represents ideal Curie-Weiss behaviour.

## AC susceptibility

The spin dynamics of the underlying freezing transitions were probed by AC susceptibility measurements using the ACMS II option for the PPMS. Samples were cooled in zero field and the response to an alternating field with a $H_{AC}$ = 5 Oe amplitude was recorded. The frequency $\nu$ of the AC field was varied between 100 and 10 000 Hz. The temperature scans of the real part of AC susceptibility, $\chi'$, are plotted for each frequency in Figure 5, for both NaCdMn$_2$F$_7$ and NaCdNi$_2$F$_7$. In both materials, we see a systematic shift of the freezing transition $T_f$, taken as the maximum of the $\chi'$ cusp, to higher temperatures with increasing AC field frequency. This is an archetypal feature of a spin-glass transition, and in conjunction with the ZFC/FC splitting seen at the transition in DC susceptibility, this provides good evidence for a glassy ground-state in these materials. However, these frequency dependent shifts are seen near the temperature limit of our instrument, and we cannot rule out the possibility of frequency-dependent sample heating that could lead to a similar observed shift in the apparent transition temperature. A dimensionless parameter quantifying the relative shift of $T_f$ per decade of frequency, $\delta T_f = \Delta T_f/(T_f \Delta \log_{10}\nu)$ was proposed by Mydosh [50] to compare between the various classes of spin-glass-like materials. The extracted values are $\delta T_f$ = 0.018(1) and 0.033(3) for NaCdMn$_2$F$_7$ and NaCdNi$_2$F$_7$, respectively, and are in line with the order of magnitude expected for insulating spin-glasses, as was also seen in the previously-reported pyrochlore fluorides NaCdCo$_2$F$_7$ (0.010), NaCaNi$_2$F$_7$ (0.024), NaSrCo$_2$F$_7$ (0.027) and NaCaCo$_2$F$_7$ (0.029). [29–33] The dynamics of the spin-freezing process can be further modelled by the empirical Vogel-Fulcher law:

$\nu = \nu_0 exp\left(-\frac{E_a}{k_B(T_f - T_0)}\right)$. This model assumes interacting clusters of spins with the intrinsic relaxation time $\tau_0 = \nu_0^{-1}$, where an activation energy $E_a$ is needed to trigger a freezing transition into a static disordered state. Here, the *ideal glass temperature*, $T_0$, is taken as the extrapolated freezing temperature at zero frequency and is usually interpreted as an estimate of the inter-cluster interaction strength. One can extract the parameters from a linear fit on the rearranged form of the Vogel-Fulcher law, expressing the frequency-dependence of the freezing temperature $T_f = T_0 - \frac{E_a}{k_B} \frac{1}{\ln(\tau_0 \nu)}$. Due to a limited experimental frequency range, however, it is physically unreasonable to fit all three parameters independently. Hence, the intrinsic relaxation time is usually fixed to a value typical for insulating spin-glasses, $\tau_0 = 10^{-12}$ s. The extraction of parameters after fixing $\tau_0$ then yields $E_a = 1.6(2)$ meV and $T_0 = 2.4(1)$ K for NaCdNi$_2$F$_7$; and $E_a = 0.64(2)$ meV and $T_0 = 1.72(1)$ K for NaCdMn$_2$F$_7$, summarised in Table 5. In both cases, the fitted ideal glass temperature is lower than that seen as the maximum of the cusp in DC susceptibility. For comparison, values found for NaCaNi$_2$F$_7$ and NaSrMn$_2$F$_7$ were $E_a = 1.4(2)$ meV and 0.83(4) meV; and $T_0 = 3.0(1)$ K and 2.2(7) K, respectively. [32,33] In Figure 6, we directly compare our fits, with the y-axis normalized by the respective ideal glass temperatures, $T_0$.

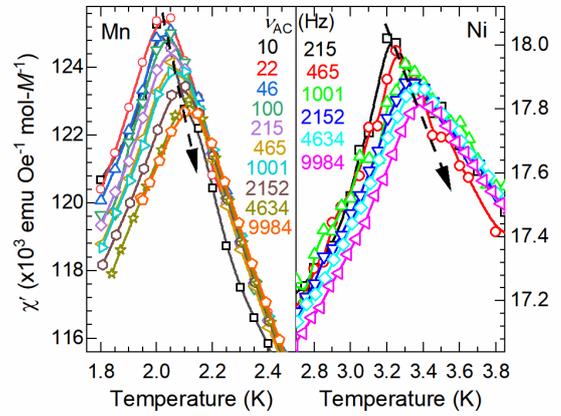

Figure 5. AC-susceptibility data, $\chi'$, for (left) NaCdMn$_2$F$_7$, and (right) NaCdNi$_2$F$_7$. The arrows highlight the shift to higher temperatures of the $\chi'$ peak with increasing frequency.

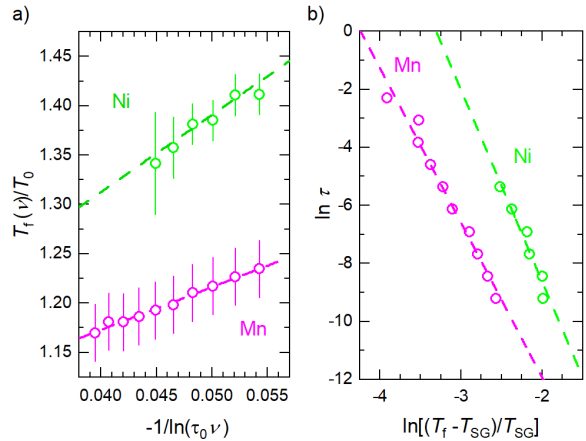

Figure 6. a) Vogel-Fulcher analysis of the frequency-dependent shift of the freezing transition, normalised to the zero-frequency freezing temperature for NaCdNi$_2$F$_7$ and NaCdMn$_2$F$_7$. b) Dynamical scaling fit for NaCdNi$_2$F$_7$ and NaCdMn$_2$F$_7$.

Table 5: Fitted ACMS parameters using the Vogel-Fulcher model and the theory of dynamical scaling. Comparison of our data with the previously-studied AA'B$_2$F$_7$ pyrochlore fluorides. *Values were not reported explicitly in [33], but instead were extracted from the Figure 8 digitized. **Dynamical scaling was not reported in [31], but the data were revisited and reanalyzed.

| Compound | Fit of $T_f(v)$ vs. log(v) | | Vogel-Fulcher analysis | | Dynamical scaling | | References |
| --- | --- | --- | --- | --- | --- | --- | --- |
| | $\delta T_f$ | $T_0$ (K) | $T_0$ (K) | $E_a$ [meV] | $\tau_0$ (s) | $zv'$ | |
| NaCdNi$_2$F$_7$ | 0.033(3) | 2.98(4) | 2.4(1) | 1.6(2) | 3.3(2) ×10$^{-10}$ | 6.6(8) | This work |
| NaCaNi$_2$F$_7$ | 0.024 | ----- | 3.0(1) | 1.4(2) | ----- | ----- | [32] |
| NaCdMn$_2$F$_7$ | 0.018(1) | 1.97(1) | 1.72(1) | 0.64(2) | 1.6(1) × 10$^{-10}$ | 5.3(3) | This work |
| NaSrMn$_2$F$_7$ | *0.018(2) | ----- | 2.2(7) | 0.83(4) | ----- | ----- | [33] |
| NaCdCo$_2$F$_7$ | 0.010(1) | **3.91(1) | 3.7(1) | 0.64(6) | **5.3(2) × 10$^{-10}$ | **4.1(4) | [31] |
| NaCaCo$_2$F$_7$ | 0.029 | ----- | 2.17 | 1.0 | ----- | ----- | [29] |
| NaSrCo$_2$F$_7$ | 0.027 | ----- | 2.6(1) | 1.3(1) | ----- | ----- | [30] |
| NaCaFe$_2$F$_7$ | *0.051(3) | ----- | 2.8(4) | 2.9(4) | ----- | ----- | [33] |
| NaSrFe$_2$F$_7$ | *0.057(3) | ----- | 2.7(2) | 2.8(1) | ----- | ----- | [33] |

Another approach to model the relaxation dynamics during the spin-freezing transition is to use the results of the standard theory of dynamical scaling near $T_f$, which gives the power-law divergence $\tau = \tau_0 \left(\frac{T_f - T_{SG}}{T_{SG}}\right)^{-zv'}$. [50,51] Here, $T_f = T_f(v)$ is the frequency-dependent maximum of the $\chi'$ cusp, $T_{SG}$ should correspond to the extrapolated value of $T_f$ as $v \rightarrow 0$, $z$ is the dynamical critical exponent, and $v'$ is the critical exponent of the correlation length. In spin glasses, the values of the product $zv'$ typically lie between 4 and 12, whereas in conventional phase transitions $zv'$ is around 2. [50] These parameters can be extracted after rearranging the power-law divergence as $\ln(\tau) = \ln(\tau_0) - zv' \ln\left(\frac{T_f - T_{SG}}{T_{SG}}\right)$ and performing a linear fit of $\ln(\tau)$ versus $\ln\left(\frac{T_f - T_{SG}}{T_{SG}}\right)$, where $\tau = v^{-1}$. The critical exponents and intrinsic relaxation times extracted from the fits are $zv' = 6.8(8)$ and 5.3(3); and $\tau_0 = 3.3(2) \times 10^{-10}$ s and $1.6(1) \times 10^{-10}$ s for NaCdNi$_2$F$_7$ and NaCdMn$_2$F$_7$, respectively. Here, $T_{SG}$ was taken as the linear extrapolation of $T_f(v)$ vs. $\ln(v)$ to zero frequency – 2.98(4) K and 1.97(1) K for NaCdNi$_2$F$_7$ and NaCdMn$_2$F$_7$, respectively. Both critical exponents lie well in the $zv' \in [4, 12]$ region expected for spin-glasses. However, the extracted relaxation times suggest two orders of magnitude slower relaxation dynamics than the estimated value $\tau_0 = 10^{-12}$ s used in the Vogel-Fulcher fits as a fixed parameter.

*Specific heat*

The magnetic ground-states of NaCdNi$_2$F$_7$ and NaCdMn$_2$F$_7$ were further probed by specific heat measurements. For the estimation of phonon contribution to the specific heat, allowing a subtraction to separate the magnetic component, a non-magnetic analogue NaCdZn$_2$F$_7$ was used (after slight individual scaling necessary due to differing unit cell volume). The temperature dependences of the specific heat are shown in Figure 7. Noting that the studied materials are insulating, no additional contribution to the specific heat from conducting electrons is present, allowing us to estimate the magnetic specific heat as $C_{mag} = C_p - C_{lattice} \approx C_{p[NaCdM2F7]} - C_{p[NaCdZn2F7]}$ ($M$ = Ni, Mn). In both materials, we see a broad peak in the magnetic specific heat below 20 K, a typical feature seen in spin-glass materials as spin-correlations build. A sharp peak can be seen roughly at $T_f$ on the $C_{mag}/T$ vs. $T$ scale, with no additional broadening after applying an external field up to 9 T. Below $T_f$, the materials exhibit a power-law decay of the magnetic specific heat, which was fitted in the $T < T_f$ region to a simple power function in the form $C_{mag} = AT^\alpha$. The extracted parameters are $A = 0.110(1)$ J mol-Ni$^{-1}$ K$^{-1-\alpha}$ and $\alpha = 1.91(1)$ for NaCdNi$_2$F$_7$, as well as $A = 0.79(1)$ mol-Mn$^{-1}$ K$^{-1-\alpha}$ and $\alpha = 1.62(3)$ for NaCdMn$_2$F$_7$. Usually, a linear dependence of the magnetic specific heat ($\alpha = 1$) is expected below $T_f$ in canonical (metallic) spin-glasses, arising from a statistical distribution of localized tunnelling levels linear in energy. [52] In the case of insulating spin-glass materials, a power-law dependence is usually observed, with the exponent typically close to $\alpha \sim 2$, as was seen in the isostructural analog NaCaNi$_2$F$_7$ ($\alpha = 2.2$). [40]

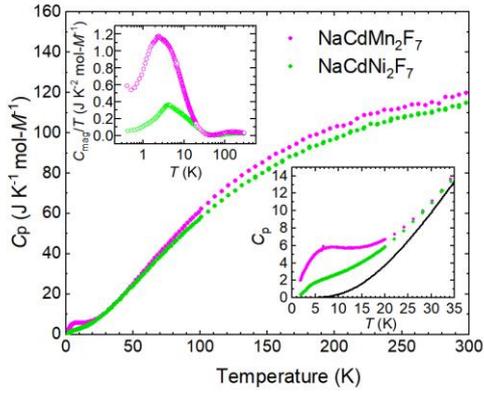

Figure 7. Heat capacity vs. temperature for NaCdMn$_2$F$_7$ and NaCdNi$_2$F$_7$. The solid line of the lower inset indicates the heat capacity of the non-magnetic analogue NaCdZn$_2$F$_7$. The upper inset shows the extracted $C_{mag}/T$ after subtraction of the phonon contribution.

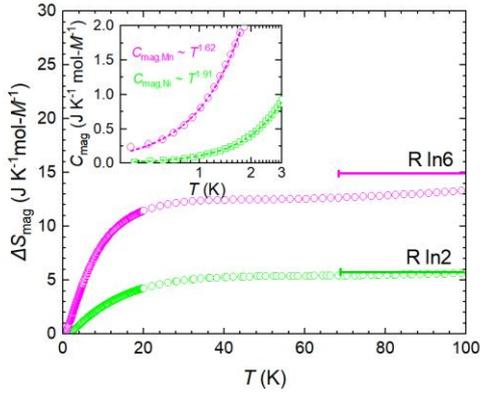

Figure 8. The extracted magnetic entropy vs. temperature for NaCdMn$_2$F$_7$ and NaCdNi$_2$F$_7$, showing saturation above ~30 K. The inset shows the low-temperature magnetic contribution to specific heat, with fits to power-law behaviour indicated by the dashed lines.

The magnetic entropy, $\Delta S_{mag}(T)$ was extracted by integration of the $C_{mag}/T$ vs. $T$ data and is shown in Figure 8. In both materials, the magnetic entropy saturates at about 50 K. In NaCdMn$_2$F$_7$ the saturation value reaches about 85% of the high-temperature Heisenberg limit expected for an $S = 5/2$ Mn$^{2+}$ ion, $S_{mag} = R \ln(2S+1) = R\ln(6)$. In contrast, the magnetic entropy of NaSrMn$_2$F$_7$ was seen to begin saturating higher than Rln(6) at 70 K. [33] In NaCdNi$_2$F$_7$, the magnetic entropy surprisingly saturates at a very low value of Rln(2), significantly short of the Heisenberg value of Rln(3) expected for a $S = 1$ Ni$^{2+}$ ion. Both materials thus suggest a large residual entropy at 0.4 K, indicating continued dynamics at low temperature. This was also seen in NaCaNi$_2$F$_7$, where the magnetic entropy extracted from specific heat measured down to 1.8 K also saturated at Rln(2) [32], although employing a dilution fridge down to 100 mK later suggested a saturation of magnetic entropy at about 84% of Rln(3), with residual entropy of 0.176R at 100 mK. [40]

## Conclusions

We report a successful single-crystal growth of the $S = 1$ pyrochlore antiferromagnet NaCdNi$_2$F$_7$ and the $S = 5/2$ defect-fluorite antiferromagnet NaCdMn$_2$F$_7$, on which we performed structural, magnetic and thermodynamic studies.

Structural analysis confirms the expected pyrochlore structure of NaCdNi$_2$F$_7$ (cubic, SG: $Fd$-$3m$, $a$ = 10.2245(10) Å), with SCXRD refinement leading to a formula Na$_{0.892}$Cd$_{1.108}$Ni$_2$F$_{6.982}$. The $A$-site cations (Na/Cd) are fully disordered, with a ~11% preference for Na$^+$ deficiency and Cd$^{2+}$ excess. NaCdMn$_2$F$_7$ crystallizes in the defect-fluorite structure (cubic, SG: Fm-3m, $a$ = 5.2461(5) Å), as the Mn$^{2+}$ ion is too similar to the average $A$-site ionic size to drive the cationic site order of the stable pyrochlore structure. For NaCdMn$_2$F$_7$, the ratio of $A$-site and $B$-site ionic radii $RR = R_A/R_B = 1.37$ lies below the lower boundary for pyrochlore structure stability [8]. Instead, Na$^+$/Cd$^{2+}$/Mn$^{2+}$ ions are fully disordered on the fluorite $A$-site with 0.25-0.25-0.5 occupancy, while the F site contains a disordered vacancy due to the 7/8 occupation in the defect fluorite structure.

Despite rather strong antiferromagnetic interactions between $Ni^{2+}$ moments in $NaCdNi_2F_7$ ($\theta_{CW}$ = -91.2(5) K), no transition is seen until a glassy freezing transition at $T_f$ = 3.2 K, indicating a large magnetic frustration ($f$ = 28.5) in this $S$ = 1 pyrochlore antiferromagnet. A spin-glass-like ground-state is noted via ZFC/FC splitting of χ at $T_f$ in a small field (100 Oe), reinforced by the frequency-dependent shift of the χ′ cusp maximum, $T_f(\nu)$, to a higher temperature. This shift was well parametrized by the empirical Vogel-Fulcher law as well as the theory of dynamical scaling, with fitted parameters typical of insulating spin-glasses, although further spin-glass characterisation measurements, such as neutron or muon spectroscopy or thermoremanent magnetisation, and suitable modelling, are necessary to conclusively label these compounds as ideal spin-glasses. Magnetic entropy of only ~ Rln(2) is recovered at 100 K, in contrast to the expected Heisenberg value for a $S$ = 1 system (R ln(3)), indicating large residual entropy at low temperature consistent with a ground-state with continued dynamics. This was also seen in the isostructural analogue $NaCaNi_2F_7$. [32,39,40]

$NaCdMn_2F_7$ shows weaker antiferromagnetic interactions between $Mn^{2+}$ spins ($\theta_{CW}$ = -42.4(4) K) compared to $NaSrMn_2F_7$ ($\theta_{CW}$ = -89.72 K), as the Mn-F distance in the Mn-F-Mn superexchange pathway is larger in the defect-fluorite structure compared to the $NaSrMn_2F_7$ pyrochlore. Although the FCC lattice offers weaker geometrical frustration in 3D than the pyrochlore lattice, no magnetic transition is seen until a spin-glass-like transition at $T_f$ = 2.0 K. This is much lower than the mean-field interaction strength, yielding a large frustration index of $f \sim 21$. AC magnetic susceptibility and specific heat measurements also confirmed the glassy nature of this transition.

The glassy ground-states uncovered in $NaCdNi_2F_7$ and $NaCdMn_2F_7$ are in line with all the previously-studied $AA'B_2F_7$-type pyrochlore fluorides. [29–33] The spin freezing is invoked by random local variations in magnetic bond lengths and angles – magnetic bond disorder. This comes as a result of the chemical disorder of the non-magnetic ions – the random mixing of Na/Cd on the pyrochlore $A$-site in $NaCdNi_2F_7$, as well as random mixing of Na/Cd/Mn ions in the defect-fluorite $A$-site in $NaCdMn_2F_7$. Monte Carlo simulations on the classical Heisenberg antiferromagnet with nearest neighbor exchange on the pyrochlore lattice reveal that the spin glass transition temperature correlates with the disorder strength as $k_B T_f = \sqrt{\frac{8}{3}} \Delta$, where Δ is the amplitude of small variations in the exchange coupling constant $J_{ij}$. [26]

Geometric magnetic frustration remains understudied in fluoride compounds, when compared to their oxide counterparts. Expanding the studies of the 2-2-7 fluorides past the border of pyrochlore structure stability provides an important comparison with the heavy rare-earth zirconate and hafnate oxides. [53–57] This initial characterisation paves the way for a more detailed investigation into the role of disorder in the geometrically frustrated pyrochlore fluoride magnets – specifically at this boundary between pyrochlore and defect fluorite structures.

*Acknowledgements*

This work was supported by the Czech Ministry of Education, Youth and Sports (MŠMT, project no. LUABA24056), the Czech Science Foundation (project no. 19-21575Y) and the Charles University (GA UK project no. 48924). The preparation, characterization and measurement of bulk physical properties were performed in MGML (https://mgml.eu), which is supported within the program of Czech Research Infrastructures (project no. LM2023065). The authors additionally thank Gaël Bastien and Milan Klicpera for useful discussions.